\documentstyle[preprint,prc,aps,epsfig]{revtex}
\def\approxlt{\kern0.35em\raise0.45ex\hbox{$<$}\kern-0.66em\lower0.5ex
  \hbox{$\scriptstyle\sim$}\kern0.35em}
\def\approxgt{\kern 0.35em\raise 0.45ex\hbox{$>$}\kern-0.75em\lower0.5ex
  \hbox{$\scriptstyle\sim$}\kern0.35em}
\def\tildew{w\llap{\lower4.5pt\hbox{$\scriptstyle\sim${}}}}
\def\ds{\displaystyle\strut}

\def\ksub0y{k_0^{\phantom{|}}\hspace*{0.02cm} y}
\def\k0rho{k_0^{\phantom{|}}\hspace*{0.02cm} \rho}
\def\usub0{u_0^{\phantom{|}}}
\def\usubk{u_k^{\phantom{|}}}
\def\qsub0{q_0^{\phantom{|}}}
\def\qsubk{q_k^{\phantom{|}}}
\def\ygt{y_{>}^{\phantom{|}}}
\def\ylt{y_{<}^{\phantom{|}}}

\begin{document}

\title{Boundary Conditions for Three-Body \\
Scattering in Configuration Space}
\author{G. L. Payne}
\address{Department of Physics and Astronomy,
The University of Iowa, Iowa City, Iowa~~52242}
\author{W. Gl\"ockle}
\address{Institut f\"ur Theoretische Physik II,
Ruhr-Universit\"at Bochum, Germany}
\author{J. L. Friar}
\address{Theoretical Division, Los Alamos National 
Laboratory, Los Alamos, New Mexico 87545}

\date{August 1999}

\maketitle

\begin{abstract}
The asymptotic behavior of three-body scattering wave functions in
configuration space is studied by considering a model equation that has
the same asymptotic form as the Faddeev equations. Boundary conditions for
the wave function are derived, and their validity is verified by numerical 
calculations. It is shown that these boundary conditions for the partial
differential equation can be used to obtain accurate numerical solutions
for the wave function.

\end{abstract}

\section{Introduction}

In a previous paper\cite{Glo92} (hereafter referred to as I), we studied the
asymptotic form of a three-body wave function, which results in the propagation
of three free particles from various types of sources. The goal was to determine
the boundary conditions appropriate for the three-body scattering equations in
configuration space. In particular, we want to establish the values of
\[
\rho={1\over2}\,\sqrt{\sum\nolimits_{i=1}^3({\bf x}_i-{\bf x}_{\rm cm})^2}
\]
for which the leading asymptotic form of an outgoing wave would be valid and to
investigate the correction terms to that form. For a complete discussion of
three-body scattering in configuration space see Ref.\ \cite{book}, which also
contains references to earlier work on this problem.

In I two sources were studied. The first was a localized source
corresponding to the elastic-scattering driving term in the three-body Faddeev
equation and is determined by the overlap of a two-body force in one pair and a
two-body bound-state wave function in another pair. The other was an extended
model source that mimics the real source term in the Faddeev equation including
the breakup process in the Faddeev amplitude.  The latter source reaches far out
in the distance $y$ between one particle and the center of mass of the other two
particles; it decreases only as $O(y^{-3/2})$.  However, the presence of the
pair interaction limits the extent in the distance $x$ between the other two
particles.  As expected in the case of the extended source, the leading form is
reached only at a much larger radius than the localized source, specifically
when $x$ is small and $y$ is large.

By inverting the free propagator, one can determine the propagating wave
function using a partial differential equation with the given source terms. We
established suitable boundary conditions that could be used to solve this
problem efficiently. A matching radius of about $100\rm\,fm$ was found to be
sufficient.

In this article we extend our previous study of the extended source to allow one
pair interaction to be present while the three particles propagate from the
given sources, which is exactly what happens in the Faddeev formulation.  We use
the same notation as in I\null.  In Sec.~II we evaluate the three-body Green's
function including one pair interaction, apply it to the extended source, and
study the asymptotic behavior in the two- and three-body fragmentation channels.
In Sec.~III we solve the related partial differential equation as an exercise
for applying that technique (in a forthcoming article) to the Faddeev equation
itself. Finally, we conclude in Sec.~IV.

\section{Three-Body Propagation From Given Sources \\
 With A Pair Interaction} 

In order to avoid unnecessary complications, we restricted our considerations
in I to three identical bosons interacting by spin-independent s-wave pairwise
interactions in a state with total angular momentum of zero. For this case the
Faddeev equation for the channel with incident wave $\Phi$,
\begin{equation}
\Psi= \Phi+G(E)VP \Psi\ , \label{add-1}
\end{equation}
where $P=P^++P^-$ is the permutation operator, reads in explicit notation as
\begin{equation}
\psi(x,y)=\phi(x,y)+\int_0^\infty dx'\int_0^\infty dy'\,g(x,y;x',y')
\,Q(x',y')\ .\label{add-2}
\end{equation}
We have introduced in Eq.\ (\ref{add-2})
the reduced Faddeev amplitude $\psi(x,y)$ and the corresponding reduced
Green's function $g(x,y;x',y')$. The coordinates $x$ and $y$
are the standard Jacobi variables

\begin{eqnarray}
{\bf x} &\equiv& {\bf x_1} = {\bf r_2-r_3} \label{xeqn} \\
{\bf y} &\equiv& {\bf y_1} = {\bf r_1}-\textstyle{1\over2}\,({\bf r_2+r_3})
\label{yeqn}
\end{eqnarray}
expressed in terms of the individual position vectors. 

The source term in Eq. (\ref{add-2}) is given by 
\begin{equation}
Q(x,y)=V(x)\int_{-1}^1d\mu\,{xy\over x_2y_2}\,\psi(x_2,y_2)\ , \label{add-3}
\end{equation}
where $x_2$, $y_2$, and $\mu$ in Eq.\ (\ref{add-3}) result from (\ref{xeqn}) 
and (\ref{yeqn}) by the cyclical
permutations, and are explicitly given by
\begin{eqnarray}
x_2 &=& \sqrt{\textstyle{1\over4}\,x^2+y^2+xy\mu} \label{x2eqn} \\
y_2 &=& \sqrt{\textstyle{9\over16}\,x^2+\textstyle{1\over4}\,y^2
-\textstyle{3\over4}\,xy\mu}\ , \label{y2eqn}
\end{eqnarray}
where $\mu$ is the cosine of the angle between $\bf x$ and $\bf y$. The total
wave function is given by the sum of the three Faddeev amplitudes. In Eq.\ 
(\ref{add-1}) $\Psi$ is one of the Faddeev amplitudes, the other two are
generated by cyclical permutations of the particles, $P^+ \Psi$ and
$P^- \Psi$, and they appear in the source term as $P \Psi$. The pair
interaction is $V(x)$ and this interaction also occurs in the Green's function
\[
G(E)={1\over E-(H_0+V)+i\varepsilon}\,.
\]
We refer to Ref.\ \cite{Globook} for the general background and details on the
notation.

The source term has a short-range component arising from the elastic-scattering
piece of the Faddeev amplitude, and a long-range component from the breakup
piece of the Faddeev amplitude. In I we studied the effects of both 
components; however, for the long-range component we used a model source term.
Using the asymptotic form of the Faddeev amplitude derived by the stationary
phase approximation, one finds \cite{book} that the asymptotic form of the
source term for three equal mass particles with total energy $E$ is
\begin{equation}
Q(x,y) \rightarrow C\,V(x)\,{x\,e^{i\sqrt{4/3}\,\ksub0y}\over y^{3/2}}\,, 
\label{qasy}
\end{equation}
where $k_0^2=mE/\hbar^2$ and the constant $C$ is given by the magnitude of the
wave function in the asymptotic region.
Therefore, to study the effects of this long-range behavior,
we used the model source term
\begin{equation}
Q_{\rm Model}(x,y)=V(x)\,{xy\,e^{i\sqrt{4/3}\,\ksub0y}\over(y+y_0)^{5/2}}\,,
\label{qmodel}
\end{equation}
with $y_0=2\rm\,fm$. This source term has the same asymptotic form as 
(\ref{qasy}), and we have set $C$ to be unity for convenience.

In I we neglected the final-state interaction between one pair in the 
propagator.  As required by the Faddeev scheme, this will now be included.
We will study the second term in Eq.\ (\ref{add-2}) with the source
term replaced by the model source. Thus, we write
\begin{equation}
F(x,y)=\int_0^\infty dx'\int_0^\infty dy'g(x,y;x',y')\,Q_{\rm Model}(x',y')\:.
\label{II-1}
\end{equation}
To simplify the numerical calculations we follow the procedure used in I, and
use the Bargmann two-body potential
\[
V(x)=-{V_0\,e^{-\lambda x}\over (1+\beta\,e^{-\lambda x})^2}\,,
\]
where
\[
V_0=2\beta\biggl({\hbar^2\lambda^2\over M}\biggr)\,.
\]
The bound-state wave function for this potential is given by
\[
u_d(x)=\sqrt{2\kappa\beta}{(1-e^{-\lambda x})\over (1+\beta e^{-\lambda x})}
e^{-\kappa x}\,,
\]
where $\kappa$ is the bound-state wave number for a two-body state with the 
energy $\epsilon=-\hbar^2\kappa^2/m$ and
\[
\beta={\lambda+2\kappa\over \lambda-2\kappa}\,.
\]
For our model calculations we use the values $\kappa=0.2316\rm\,fm^{-1}$ and
$\lambda=0.7\rm\,fm^{-1}$. 
We also need the two-body scattering states $\usubk(x)$ for this potential. 
They are given by
\[
\usubk(x)={1\over2i}\,\Big\{e^{i\delta(k)}e^{ikx}h(k,x)-e^{-i\delta(k)}
e^{-ikx}h^*(k,x)\Big\}\:,
\]
with
\[
h(k,x)={1+\left( {\ds2k-i\lambda\over\ds2k+i\lambda} \right) \,
\beta\, e^{-\lambda x} \over
1+\beta\, e^{-\lambda x}}\:,
\]
and
\[
e^{i\delta(k)}=\sqrt{\ds2k+i\lambda\over\ds2k-i\lambda}\,
\sqrt{\ds k+i\kappa\over\ds k-i\kappa}\:.
\]
In addition, we use the arbitrary but fixed laboratory energy of the incident
particle of $E_{\rm lab}=14\rm\,MeV$ and $\hbar^2\!/m=41.47\rm\,MeV\cdot fm$
which corresponds to the case for three nucleons. For this case
$k_0=0.41403\rm\,fm^{-1}$. Henceforth, we set $\hbar=1$.

For three equal-mass particles with mass $m$ and total energy $E$,
the three-body Green's function $g(x,y;x',y')$ for the case with one pair
interaction has the well-known form
\begin{eqnarray}
g(x,y;x',y') &=& u_d(x)\biggl(-{4m\over3}\,e^{i\qsub0 y_{>}}\,{\sin \qsub0 \ylt\over 
\qsub0}\biggr)u_d(x')  \nonumber \\
   & & {} +{2\over\pi}\int_0^\infty dk\,\usubk(x)\biggl(-{4m\over3}\,
{e^{i\qsubk \ygt}\,\sin \qsubk \ylt\over \qsubk}\biggr)\usubk(x')\:, \label{II-2}
\end{eqnarray}
where $\qsub0=\sqrt{4m/3(E-\epsilon)}$, $\qsubk=\sqrt{4/3(k_0^2-k^2)}$, and $k_0^2=
mE$. Obviously the propagation from the source can now proceed not only into the
unbound states but also into the deuteron channel. Moreover, the two-body
scattering states $\usubk(x)$ include the interaction $V(x)$.  It is the purpose of
this paper to study the additional effects of $V(x)$ in the propagator, in
contrast to I where only the free propagator was considered.  The numerical
evaluation of the second part in (\ref{II-2}) containing $\usubk(x)$ requires some
explanation.  We were not able to find an analytical expression such as we found
for $g_0$ in I, and had to perform the $k$-integral numerically.  Clearly at the
upper end of the integral the integrand has rapid oscillations that make the
integral difficult to evaluate numerically. Since $\usubk(x)$ approaches $\sin
kx$ as $k\to\infty$, it appears natural to subtract the free propagator
$g_0(x,y;x',y')$ and add it back in a separate term.  Then for $k\to\infty$ the
integrand has a stronger fall off and, moreover, the path of integration can be
rotated into the complex plane, similar to the procedure used in Ref.\
\cite{Glo88}.

Let us start with the propagation in the deuteron channel:
\begin{eqnarray}
F_d(x,y) &\equiv & -{4m\over3}\,u_d(x)\int_0^\infty dy'\,{ e^{i\qsub0 \ygt}\,
\sin \qsub0 \ylt\over \qsub0}\int_0^\infty dx'u_d(x')\,Q_{\rm Model}(x',y') \nonumber \\
 & = & -{4m\over3}\,u_d(x)e^{i\qsub0 y}\int_0^\infty dy'\,{\sin \qsub0 y'\over
\qsub0}\int_0^\infty dx'u_d(x')\,Q_{\rm Model}(x',y')\nonumber \\
 & & -{4m\over3}\,u_d(x)\int_y^\infty dy'\,{\sin \qsub0(y-y')\over
\qsub0}\int_0^\infty dx'u_d(x')\,Q_{\rm Model}(x',y')\:. \label{II-3}
\end{eqnarray}
We see that this consists of a flux-conserving term
\begin{equation}
F^{asy}_d(x,y)\, = u_d(x)e^{i\qsub0 y}f_d\:, \label{II-4}
\end{equation}
with
\[
f_d=-{4m\over3}\int_0^\infty dy'\,{\sin \qsub0 y'\over \qsub0}\int_0^\infty
dx'u_d(x')\,Q_{\rm Model}(x',y')\:,
\]
and a correction term
\begin{equation}
F_d^{\,\rm corr}(x,y)\equiv-{4m\over3}\,u_d(x)\int_y^\infty dy'\,
{\sin \qsub0(y-y')\over \qsub0}\int_0^\infty dx'u_d(x')\,Q_{\rm Model}(x',y')\:. 
\label{II-5}
\end{equation}

Inserting Eq.\ (\ref{qmodel}) into Eq.\ (\ref{II-5}) and performing one partial
integration in the $y'$ variable, one arrives easily at the following asymptotic
form
\begin{equation}
F_d^{\,\rm corr}(x,y)\,{\lower7pt\hbox{$\displaystyle\longrightarrow$}
\atop\scriptstyle{y\rightarrow\infty}}\,-{e^{i\sqrt{4/3}\,\ksub0y}\over y^{3/2}}\,
u_d(x)\,{1\over\epsilon}\int_0^\infty dx'u_d(x')x'V(x')\:. \label{II-7}
\end{equation}
Clearly the long-range source behavior carries over into a corresponding 
long-range correction term in the deuteron channel.  We rewrite Eq.\ 
(\ref{II-5}) in the form
\begin{equation}
F_d^{\,\rm corr}(x,y) = {e^{i\sqrt{4/3}\,\ksub0y}\over y^{3/2}}\,
u_d(x)\,f_d^{\,\rm corr}(y)\:,
\end{equation}
and show in Fig.~1 the behavior of $f_d^{\,\rm corr}(y)$ as it approaches its
asymptotic value given by
\begin{equation}
f_d^{\,\rm corr}(y)\,{\lower7pt\hbox{$\displaystyle\longrightarrow$}
\atop\scriptstyle{y\rightarrow\infty}}\,-{1\over\epsilon}
\int_0^\infty dx'u_d(x')x'V(x')\:. \label{II-7p}
\end{equation}
To illustrate the convergence, we have normalized the plot to the asymptotic
value $f_d^{\,\rm corr}(\infty) = -23.325$\,fm$^{3/2}$.

To illustrate the error in the elastic term that results from matching to the
asymptotic boundary conditions at a finite distance, we plot in Fig.~2 the
absolute value of $f_d(y)\equiv F_d(x,y)/u_d(x)$ and its asymptotic form given
by Eq.\ (\ref{II-4}) and Eq.\ (\ref{II-7}). The difference is less than 2\% for
$y$ greater than 50\,fm and less than 1\% for $y$ greater than 75\,fm.

The propagation into the unbound states $\usubk(x)$ is more complicated, since
\begin{equation}
F_{\rm scat}(x,y)\equiv-{4m\over3}\,{2\over\pi}\int_0^\infty dk\,\usubk(x)
\int_0^\infty dy'{e^{i\qsubk \ygt}\,\sin \qsubk \ylt\over \qsubk}
\int_0^\infty dx'\,\usubk(x')\,Q_{\rm Model}(x',y') \label{II-8}
\end{equation}
can be written in the form
\begin{eqnarray}
F_{\rm scat}(x,y) &=& -{4m\over3}\,{2\over\pi}\int_0^{k_0}dk\,\usubk(x)
e^{iqy}\,\int_0^\infty dy'\,{\sin \qsubk y'\over \qsubk}\int_0^\infty dx'
\usubk(x')\,Q_{\rm Model}(x',y')\nonumber \\
& & -{4m\over3}\,{2\over\pi}\int_0^{k_0}dk\,\usubk(x)
\int_y^\infty dy'\,{\sin \qsubk(y-y')\over \qsubk}\int_0^\infty dx'
\usubk(x')\,Q_{\rm Model}(x',y') \nonumber \\
& & -{4m\over3}\,{2\over\pi}\int_{k_0}^\infty dk\,\usubk(x)
e^{-Ky}\,\int_0^y dy'\,{{\rm sinh} \ Ky'\over K}\int_0^\infty dx'
\usubk(x')\,Q_{\rm Model}(x',y')\nonumber \\
& & -{4m\over3}\,{2\over\pi}\int_{k_0}^\infty dk\,\usubk(x)\,
{{\rm sinh} \ Ky\over K}\int_y^\infty dy'e^{-Ky'}\int_0^\infty dx'
\usubk(x')\,Q_{\rm Model}(x',y')\label{II-9} \\
&\equiv& F_{\rm scat}^{(1)}(x,y)+F_{\rm scat}^{(2)}(x,y)+F_{\rm scat}^{(3)}(x,y)
+F_{\rm scat}^{(4)}(x,y)\:. \label{II-10}
\end{eqnarray}
In the third and fourth terms $K\equiv\sqrt{4/3}\,\sqrt{k^2-k_0^2}$.

Let us first examine the asymptotic behavior for fixed $x$ and $y$ approaching
infinity.  One has
\begin{equation}
F_{\rm scat}^{(1)}(x,y)=\int_0^{k_0}dk\,\usubk(x)e^{i\qsubk y}\,T(k)
\label{II-11}
\end{equation}
with
\begin{equation}
T(k)=-{4m\over3}\,{2\over\pi}\int_0^\infty dy'\,{\sin \qsubk y'\over \qsubk}
\int_0^\infty dx' \usubk(x')\,Q_{\rm Model}(x',y')\:. \label{II-12}
\end{equation}
There is no saddle-point for this case; thus, the asymptotic form arises from
the leading end-point contribution at $k=0$, which is easily evaluated to be
\begin{equation}
F_{\rm scat}^{(1)}(x,y)\big|_{k\approx0}\longrightarrow -{\sqrt{\pi}\over4}\,
3\,{^{3/4}}k_0^{3/2}\,e^{i\pi/4}\,{e^{i\sqrt{4/3}\ksub0y}\over y^{3/2}}\,
\tilde \usub0(x){\tilde T}_0\,, \label{II-13}
\end{equation}
where $\tilde \usub0(x)\equiv \usubk(x)/k|_{k=0}$
and ${\tilde T}_0 \equiv T(k) / k \mid _{k=0}$. 
We find that 
\begin{equation}
{\tilde T}_0=-{4m\over3}\,{2\over\pi}\int_0^\infty 
dy'\,{\sin\sqrt{4\over3}\,\ksub0y'\over\sqrt{4\over3}\,k_0}
\int_0^\infty dx' \tilde \usub0(x')\,Q_{\rm Model}(x',y')\:. \label{II-15}
\end{equation}
This term has the same dependence on $y$ as the correction term Eq.\
(\ref{II-7}) in the deuteron channel. Note  that ${\tilde T}_0$ is given by the
analytical expression Eq.\ (\ref{II-12}) differentiated with respect to $k$
under the integral. This is obviously true for the localized source; however,
for the extended source, one must rewrite the integral using a contour
deformation before performing the differentiation.

Let us now consider $F_{\rm scat}^{(2)}(x,y)$ in Eq.\ (\ref{II-9}) for the model
source term given by Eq.\ (\ref{qmodel})
\begin{eqnarray}
F_{\rm scat}^{(2)}(x,y) \,\longrightarrow
 -{4m\over3}\,{2\over\pi}\int_0^{k_0}
dk&\,&\usubk(x)\int_0^\infty dx'\,\usubk(x')x'V(x') \nonumber \\
& & {} \times \int_y^\infty dy'\,{\sin \qsubk(y-y')\over \qsubk}\,
{e^{i\sqrt{4/3}\,\ksub0y'}\over y'^{\,3/2}}\:. \label{II-17}
\end{eqnarray}
After one partial integration one finds
\begin{equation}   
F_{\rm scat}^{(2)}(x,y)\,
{{\lower35pt\hbox{$\displaystyle\longrightarrow$}
\atop \scriptstyle y\rightarrow\infty}
\atop\scriptstyle x\ {\rm fixed}}
-{2\over\pi}\int_0^{k_0}
dk\,\usubk(x)\,{m\over k^2}\,\int_0^\infty dx'\,\usubk(x')x'V(x')\,
{e^{i\sqrt{4/3}\,\ksub0y}\over y^{3/2}}\:. \label{II-18}
\end{equation}
The term $F_{\rm scat}^{(3)}(x,y)$, as it stands, is less obvious in its
asymptotic behavior.  The contributions to the $y'$-integral keep growing
towards the upper limit $y$. Because of the factor $\exp(-Ky)$, only
contributions from the upper end of the $y'$ integral have to be considered. 
Again by partial integration, one easily finds
\begin{eqnarray}
F_{\rm scat}^{(3)}(x,y) \,
{{\lower35pt\hbox{$\displaystyle\longrightarrow$}
\atop \scriptstyle y\rightarrow\infty}
\atop\scriptstyle x\ {\rm fixed}}
 -{4m\over3}\,{2\over\pi}\,{e^{i\sqrt{4/3}\,\ksub0y}\over y^{3/2}}
\int_{k_0}^\infty dk&\,&\usubk(x)\,{1\over2K}\,{1\over K+i\,\sqrt{4/3}\,k_0}
\nonumber \\
& & \times\int_0^\infty dx'\,\usubk(x')x'V(x')\:. \label{II-19}
\end{eqnarray}
Finally, the last piece, $F_{\rm scat}^{(4)}(x,y)$, can again be handled in 
a straightforward manner with the result
\begin{eqnarray}
F_{\rm scat}^{(4)}(x,y) \,
{{\lower35pt\hbox{$\displaystyle\longrightarrow$}
\atop \scriptstyle y\rightarrow\infty}
\atop\scriptstyle x\ {\rm fixed}}
 -{4m\over3}\,{2\over\pi}\,{e^{i\sqrt{4/3}\,\ksub0y}\over y^{3/2}}
\int_{k_0}^\infty dk&\,&\usubk(x)\,{1\over2K}\,{1\over K-i\,\sqrt{4/3}\,k_0}
\nonumber \\
\noalign{\vskip7pt}
& & \times\int_0^\infty dx'\,\usubk(x')x'V(x')\:. \label{II-20}
\end{eqnarray}
Adding equations (\ref{II-18}), (\ref{II-19}), and (\ref{II-20}) we obtain the
concise result given in Ref.\ \cite{Glo74},
\begin{equation}
F_{\rm scat}^{(2)} + F_{\rm scat}^{(3)}+F_{\rm scat}^{(4)}\longrightarrow
{2\over\pi}\, {e^{i\sqrt{4/3}\,\ksub0y}\over y^{3/2}}
\,\int_0^\infty dk\,\usubk(x)\,{m\over k^2}\,\int_0^\infty dx'\,\usubk(x')
x'V(x')\:. \label{II-21}
\end{equation}
This can be simplified by using a technique suggested by 
C. Gignoux\cite{Gignoux}. Writing the two-body Green's function in the form
\begin{equation}
g_2(x,x';z)\equiv u_d(x)\,{1\over z-\epsilon}\,u_d(x')+{2\over\pi}
\int_0^\infty dk\,\usubk(x)\,{1\over z-k^2/m}\,\usubk(x')\:, \label{II-22}
\end{equation}
the integral over $k$ occurring in Eq.\ (\ref{II-21}) can be rewritten as
\begin{equation}
-{2\over\pi}\int_0^\infty dk\,\usubk(x)\,{m\over k^2}\,\usubk(x')=
g_2(x,x';0)+u_d(x)\,{1\over\epsilon}\,u_d(x')\:. \label{II-23}
\end{equation}
Thus, we are led to the function
\begin{equation}
\tilde u(x)\equiv\int_0^\infty dx'\,g_2(x,x';0)x'V(x')\:,\label{II-24}
\end{equation}
which obeys the inhomogeneous equation
\begin{equation}
\Big[-{1\over m}\,{d^2\over dx^2}+V(x)\Big]\tilde u(x)=-xV(x)\:.\label{II-25}
\end{equation}
Using the explicit form for $g_2(x,x';0)$, one easily derives
\begin{equation}
\tilde u(x)\,
{\lower7pt\hbox{$\scriptstyle\longrightarrow$}
\atop\scriptstyle{x\rightarrow\infty}}
-\int_0^\infty dx'\,\tilde \usub0(x')x'V(x')=a\:, \label{II-26}
\end{equation}
where $a$ is the scattering length defined by $\delta(k)$ approaching $-ka$ as
$k$ goes to zero.  From Eq.\ (\ref{II-25}) and Eq.\ (\ref{II-26}) follows
\begin{equation}
\tilde u(x)=\tilde \usub0(x)-x\:.\label{II-27}
\end{equation}

We now have the concise form
\begin{eqnarray}
F_{\rm scat}^{(2)}+F_{\rm scat}^{(3)}+F_{\rm scat}^{(4)} \,
{{\lower35pt\hbox{$\displaystyle\longrightarrow$}
\atop \scriptstyle y\rightarrow\infty}
\atop\scriptstyle x\ {\rm fixed}}
{e^{i\sqrt{4/3}\,\ksub0y}\over y^{3/2}}
\Big[\tilde u(x)+u_d(x)\,{1\over\epsilon}\int_0^\infty dx'\,
u_d(x')x'V(x')\Big]\:. \label{II-28}
\end{eqnarray}
Altogether $F_{\rm scat}(x,y)$ has the asymptotic form
\begin{eqnarray}
F_{\rm scat}(x,y) \,
{{\lower35pt\hbox{$\displaystyle\longrightarrow$}
\atop \scriptstyle y\rightarrow\infty}
\atop\scriptstyle x\ {\rm fixed}}
 {e^{i\sqrt{4/3}\,\ksub0y}\over y^{3/2}}
\Big[ - \tilde \usub0(x&)&{\sqrt{\pi\over4}}\,
3\,{^{3/4}}k_0^{3/2}\,e^{i\pi/4}\,{\tilde T}_0
 +\tilde \usub0(x)-x \nonumber \\
& & +u_d(x)\,{1\over\epsilon}\int_0^\infty
dx'\,u_d(x')x'V(x')\Big] \:. \label{II-31}
\end{eqnarray}
The $x$-dependence is therefore built up of the zero-energy scattering state, a
linear term in $x$, and the two-body bound state.  The last term cancels exactly
against the correction term Eq.\ (\ref{II-7}) in the deuteron channel and the
total amplitude $F(x,y)$ behaves as\footnote{An alternate derivation of this
result is given in Section 6.3 of Ref.\ \cite{book}; however, the solution
$g(x)= -1$ of Eq.\ (2.6.19) in Ref.\ \cite{book} is not given explicitly.}
\begin{eqnarray}
F(x,y) \,
{{\lower35pt\hbox{$\displaystyle\longrightarrow$}
\atop \scriptstyle y\rightarrow\infty}
\atop\scriptstyle x\ {\rm fixed}}
u_d(x)e^{i\qsub0 y}f_d+{e^{i\sqrt{4/3}\,\ksub0y}\over y^{3/2}}
\bigg[ - \tilde \usub0(x){\sqrt{\pi\over4}}\,3\,{^{3/4}}k_0^{3/2}\,
e^{i\pi/4}\,{\tilde T}_0+\tilde \usub0(x)-x\bigg]\:. \label{II-32}
\end{eqnarray}
For $x$ outside the range of $V(x)$, the expression in the brackets in
Eq.\ (\ref{II-32}) reduces to
\begin{equation}
- (x-a){\sqrt{\pi}\over4}\,3\,{^{3/4}}k_0^{3/2}\,
e^{i\pi/4}\,{\tilde T}_0-a\:. \label{II-33}
\end{equation}

To verify the validity of the asymptotic term in Eq.\ (\ref{II-31}), we use 
Eq.\ (\ref{II-1}) to numerically evaluate $F(x,y)$ for
several values of $y$ using only the breakup component of the Green's function
given in Eq.\ (\ref{II-2}). We show the convergence to the asymptotic result by
rewriting $F_{\rm scat}(x,y)$ in the form
\begin{equation}
F_{\rm scat}(x,y) = {e^{i\sqrt{4/3}\,\ksub0y}\over y^{3/2}}\,a(x,y)\:,
\label{axy}
\end{equation}
where from Eq.\ (\ref{II-31}) $a(x,y)$ has the asymptotic form
\begin{equation}
- \tilde \usub0(x){\sqrt{\pi\over4}}\, 3\,{^{3/4}}k_0^{3/2}\,e^{i\pi/4}\,{\tilde
T}_0 +\tilde u(x)+u_d(x)\,{1\over\epsilon}\int_0^\infty
dx'\,u_d(x')x'V(x')\,. \label{ayinf}
\end{equation}
The results for several values of $y$ and the asymptotic form are shown in
Fig.~3, where one can see that $a(x,y)$ approaches its asymptotic form for
large values of $y$. To better illustrate the convergence, in Fig.~4 we plot
$a(x,y)$ versus $1/y$ for $x=3$. As $1/y$ goes to zero the plot approaches its
asymptotic value $a(3,\infty)=6.863-4.005i$ given by Eq.\ (\ref{ayinf}).

Let us now regard the asymptotic form of $F_{\rm scat}(x,y)$ for both $x$ and
$y$ going to infinity at a certain fixed angle $\theta$ in the first quadrant. 
For this case only the breakup part contributes.  Its first term $F_{\rm
scat}^{(1)}$ receives contributions from a saddle point and from the two end
points.  The result is
\begin{equation}
F_{\rm scat}(x,y)\longrightarrow{e^{i\k0rho}\over(\k0rho)^{1/2}}\,\Big[
A(\theta)+{1\over \k0rho}\, B(\theta)+\cdots\Big]\:, \label{II-34}
\end{equation}
where $\theta$ and $\rho$ are defined by $x=\rho\cos\theta,y=\sqrt{3/4}\,
\rho\sin\theta$, and 
\begin{equation}
A(\theta)=-k_0\,e^{i\pi/4}\sqrt{\pi\over2}\,\sin \theta\,
e^{i\delta(k_0\cos\theta)}T(k_0\cos\theta)\:, \label{II-35}
\end{equation}
\begin{equation}
B(\theta)={1\over2i}\,\Big[{1\over4}\,A(\theta)+A''(\theta)\Big]\:. 
\label{II-36}
\end{equation}
In the following discussion we use $(x,y)$ and $(\rho,\theta)$ interchangeably.
The end-point contributions are of $O(\rho^{-2})$ and are beyond what is 
displayed in Eq.\ (\ref{II-34}).  The relation between $B(\theta)$  and
$A(\theta)$ is the same as found for the free propagation case considered in I.
A tedious analytical study reveals that for this case the only contributions
up to order $\rho^{-3/2}$ are from $F_{\rm scat}^{(1)}$.
Thus the two leading terms in Eq.\ (\ref{II-34}) result solely from $F_{\rm
scat}^{(1)}$ in Eq.\ (\ref{II-10}). The same is of course true for the free
propagator $g_0$ studied in I\null.

It is clear that the first term alone in Eq.\ (\ref{II-34}), the flux-conserving
breakup behavior, is not a valid representation of $F_{\rm scat}$ at small
$\rho$ values.
The correction term is suppressed only by $O[(\k0rho)^{-1}]$, and depending 
upon the size of $B(\theta)$ relative to $A(\theta)$ the value of $\rho$ may
have to be very large before one can neglect the second- and higher-order terms
in Eq.\ (\ref{II-34}). To illustrate this property we numerically evaluate 
Eq.\ (\ref{II-11}) at $\rho_{\rm m}$ and $\rho_{\rm m} \pm 10$\,fm for a fixed
value of $\theta$, which is then fit to the function
\begin{equation}
\biggl[a_{\rm m} +{b_{\rm m} \over \k0rho}+{c_{\rm m} \over
(\k0rho)^2}\biggr]\, {e^{i\k0rho}\over(\k0rho)^{1/2}}\,. \label{series}
\end{equation}
The $a_{\rm m}$ for three values of $\theta$ along with the asymptotic value
$A(\theta)$ are given in Table I.

From Table I one can see that Eq.\ (\ref{series}) provides an accurate
approximation to $F_{\rm scatt}(x,y)$ at reasonable values of $\rho$. In
addition, we note that the value of $\rho_{\rm m}$ required for convergence
increases as $\theta$ increases. This feature is due to the property that large
values of $\theta$ correspond to small values of $x$, and for $x$ small one must
use Eq.\ (\ref{II-31}). From the Taylor-series expansion in $x$ of Eq.\
(\ref{II-31}) one finds that for $x$ greater than the range of the bound state
and $(x/\rho)^2$ small
\[
F_{\rm scat}(x,y) \longrightarrow A(\theta)\,{e^{i\k0rho}\over(\k0rho)^{1/2}}\,.
\]
For $\theta=80^\circ$ the value of $\rho$ must be larger than $200\,$fm for
this approximation to be valid.

\section{The Partial-Differential Equation Approach}

To test the accuracy of solving the differential form of the Faddeev equations
in configuration space we solve the partial differential equation
\begin{equation}
g^{-1}F=Q\:,\label{II-37}
\end{equation}
which has the explicit form
\begin{equation}
\left[-{1\over m}\left({\partial^2\over\partial\rho^2}+{1\over\rho}
{\partial\over\partial\rho}+{1\over\rho^2}\,{\partial^2\over\partial\theta^2}
\right)+V(\rho\cos\theta)-E\right]F(\rho,\theta)=-Q(\rho,\theta)\:,
\label{II-42}
\end{equation}
and impose boundary conditions along a quarter circle in the first $x$-$y$
quadrant at $\rho=\rho_{\rm max}$. To solve the partial differential equation,
we write
\begin{equation}
F(\rho,\theta)\equiv {e^{i\k0rho}\over{(\k0rho + \beta)^{1/2}}}\,f(\rho,
\theta)\label{II-38}
\end{equation}
and solve the resulting partial differential equation for $f(\rho,\theta)$.
The constant $\beta$ is a parameter introduced to avoid singular behavior at
the origin.
We solved this equation using the spline expansion methods described in
\cite{Portugal} for $\rho < \rho_{\rm max}$ with the boundary condition that
$F(\rho,\theta)$ have the form at $\rho_{\rm max}$ specified by Eq.\ 
(\ref{series}).

To show that this procedure can be used to obtain accurate results for the
scattering wave function, we solved the partial differential equation
numerically for various values of $\rho_{\rm max}$ and compared the results to
those obtained by numerically integrating the Green's function integral in Eq.\
(\ref{II-11}). We found that values of $\rho_{\rm max}$ on the order of 100\,fm
can be used to obtain good wave functions. Larger values of $\rho_{\rm max}$
yield more accurate solutions. In Figs.~5 and 6 we have plotted examples of
the comparisons for fixed $\rho$ values of 25\,fm and 50\,fm. One can see that
in both cases the agreement is excellent.

For the three-body scattering problem one wants to obtain accurate values of
the breakup amplitude $A(\theta)$. Since this corresponds to the amplitude of
the scattering wave function at infinity, its value cannot be obtained by
evaluating the Faddeev amplitude at large values of $\rho_{\rm max}$. In Fig.~7
we compare the ``exact'' result for $A(\theta)$ evaluated using Eq.\
(\ref{II-35}) and the integral form for $T(k)$ given in Eq.\ (\ref{II-12}) to
the values extracted from the wave function evaluated at $\rho_{\rm
max}=200\,$fm. To demonstrate again that the boundary conditions for the
partial differential equation have been treated correctly, we show the
$A(\theta)$ extracted from the numerical solution of the Faddeev equation and
the evaluation of the Green's function integral with $\rho=\rho_{\rm max}$.
While the wave function results are similar to the "exact" values, one can see
that for large values of $\theta$ they are different for the reasons discussed
in the previous section. The $A(\theta)$ obtained from the wave function at
$\rho_{\rm max}$ still exhibits the small $x$ behavior of the wave function for
$\theta$ near $90^\circ$. Thus, to obtain accurate results for $A(\theta)$ one
must use the integral expression. This was the procedure followed in Ref.\
\cite{benchmark} where it was shown that configuration-space Faddeev
calculations gave results in excellent agreement with the momentum-space
calculations.

\section{Conclusions}

Using a model source that mimics the real source term in the configuration-space
Faddeev equations, the validity of the expressions for the asymptotic behavior
of the wave function has been verified by numerically integrating the 
integral representation of the Green's
function integral. Using the asymptotic expressions as the boundary conditions
for the partial differential equation in configuration space, it is possible to
obtain an accurate solution of the scattering equation for reasonable values of
$\rho_{\rm max}$. While larger values of $\rho_{\rm max}$ are required to obtain
more accurate solutions, values on the order of 100\,fm yield good solutions. 
The breakup amplitude corresponding to small values of $x$ cannot be obtained
from the wave function evaluated at large values of $\rho$; the T-matrix
integral must be used to determine $A(\theta)$ in this region.

\acknowledgments
The work of J.\ L.\ F.\ was performed under the auspices of the U.\ S.\
Department of Energy. That of G.\ L.\ P.\ was supported in part by the U.\ S.\
Department of Energy.

\pagebreak

\newcommand{\STRUT}{\rule{0in}{3ex}}
\large

\renewcommand{\baselinestretch}{1} \tiny \normalsize
\newenvironment{Tabular}%
     {\tabcolsep 0.1in\begin{tabular}}{\end{tabular}}
\noindent
Table I: Values of $a_{\rm m}$ determined from Eq.\ (\ref{series}) for
$\theta=30^\circ$, $\theta=60^\circ$, and $\theta=80^\circ$. The values for
$\rho_{\rm m}=\infty$ were determined from Eq.\ (\ref{II-35}).

\medskip
\begin{center}
\begin{Tabular}{|r|cc|cc|cc|} \hline
\multicolumn{1}{|c}{$\rho_{\rm m}$}&
\multicolumn{2}{|c}{80$^\circ$}&
\multicolumn{2}{|c}{60$^\circ$}&
\multicolumn{2}{|c|}{30$^\circ$} \\ \hline \hline
\multicolumn{1}{|c}{}& 
\multicolumn{1}{|c}{Real}&
\multicolumn{1}{|c}{Imag}&
\multicolumn{1}{|c}{Real}&
\multicolumn{1}{|c}{Imag}&
\multicolumn{1}{|c}{Real}&
\multicolumn{1}{|c|}{Imag} \\ \cline{2-7}
   50.0{ }\STRUT &  -0.2112 & 0.1240 &  0.0595 & 0.2159 & 0.0173  & 0.0657 \\
   90.0{ } &  -0.1088 & 0.1718 &  0.0137 & 0.1921 & 0.0152  & 0.0642 \\
  140.0{ } &  -0.0737 & 0.1651 &  0.0083 & 0.1969 & 0.0155  & 0.0642 \\
  190.0{ } &  -0.0678 & 0.1609 &  0.0074 & 0.1980 & 0.0156  & 0.0642 \\
  240.0{ } &  -0.0665 & 0.1586 &  0.0070 & 0.1984 & 0.0157  & 0.0642 \\
  290.0{ } &  -0.0660 & 0.1578 &  0.0069 & 0.1986 & 0.0157  & 0.0642 \\
  340.0{ } &  -0.0659 & 0.1574 &  0.0068 & 0.1988 & 0.0157  & 0.0642 \\
 $\infty${ }{ } & -0.0657 & 0.1560  & 0.0066 & 0.1990 & 0.0157 & 0.0642 \\
\hline
\end{Tabular}
\end{center}
\renewcommand{\baselinestretch}{1.5} \tiny  \normalsize
\pagebreak

\begin{center}
{\bf FIGURE CAPTIONS}
\end{center}
\medskip

\noindent Figure~1. The behavior of $f_d^{\,\rm corr}(y)$ normalized to its
asymptotic value for $y\rightarrow\infty$.
\vskip11pt

\noindent Figure~2. Comparison of the absolute value of $f_d(y)$ to its 
asymptotic form.
\vskip11pt

\noindent Figure~3. The $x$ dependence of $a(x,y)$ defined in Eq.\ 
(\ref{axy}) for fixed values of $y$ and its asymptotic form.
\vskip11pt

\noindent Figure~4. The real and imaginary parts of $a(x,y)$ for $x=3$
plotted versus $1/y$. The triangles are the calculated values and the solid
line is a fit to a polynomial in $1/y$.
\vskip11pt

\noindent Figure~5. Comparison of the $f(\rho,\theta)$ evaluated using the
Green's function integral and the $f(\rho,\theta)$ obtained from solving the
partial differential equation for $\rho=25$\,fm.
\vskip11pt

\noindent Figure~6. Same as for Fig. 5 for $\rho=50$\,fm.
\vskip11pt

\noindent Figure~7. Comparison of the breakup amplitude, $A(\theta)$, 
evaluated using the integral form for $T(k)$ with the $A(\theta)$ extracted
from the wave function at $\rho_{max} = 200$ fm. Wave function results for
both the numerical solution of the Faddeev equation and the Green's function
integral are shown.

\pagebreak

\begin{figure}
\begin{center}
{\Large \sf FIGURE 1}
\medskip
\epsfig{file=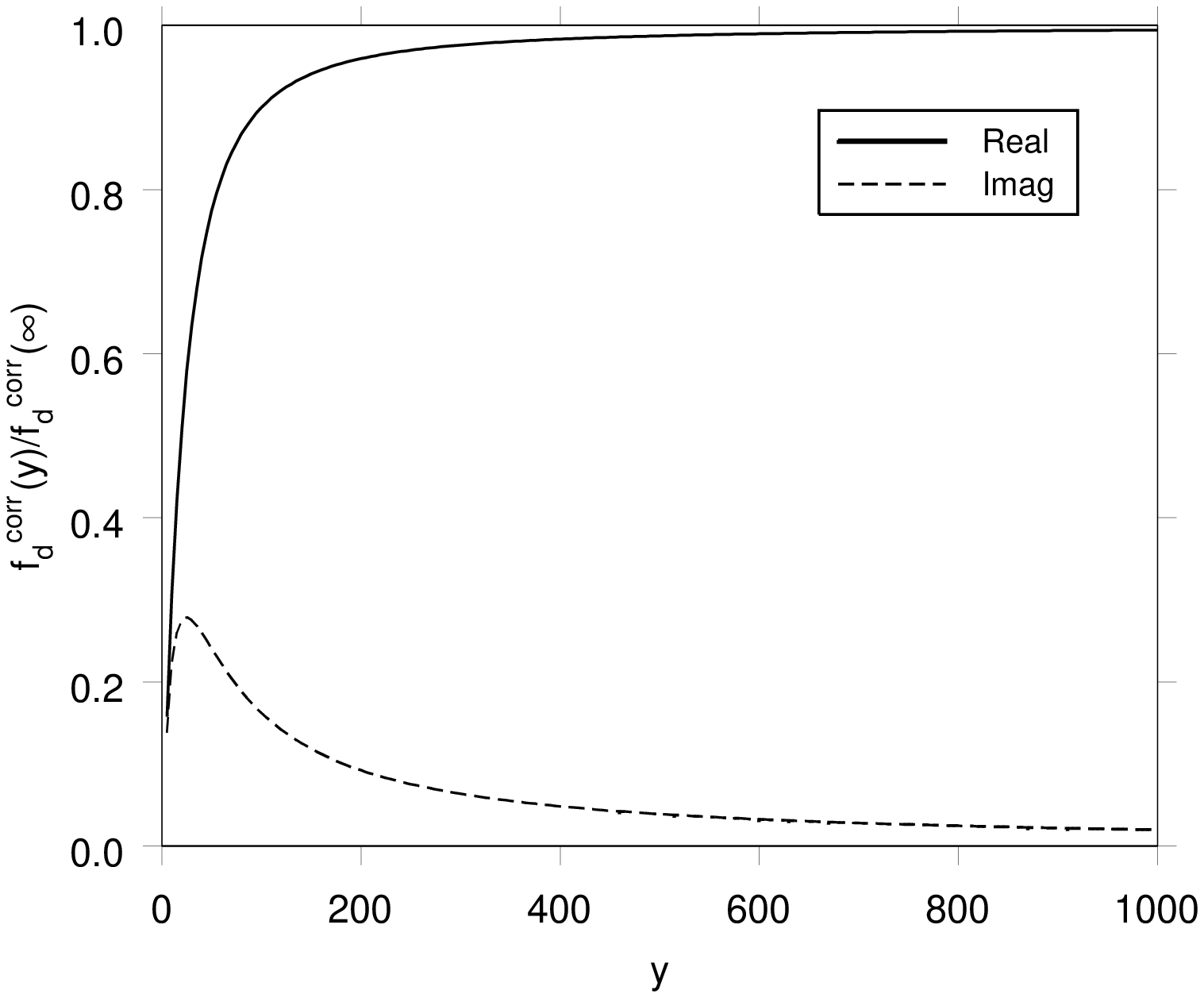}
\end{center}
\end{figure}

\pagebreak

\begin{figure}
\begin{center}
{\Large \sf FIGURE 2}
\medskip
\epsfig{file=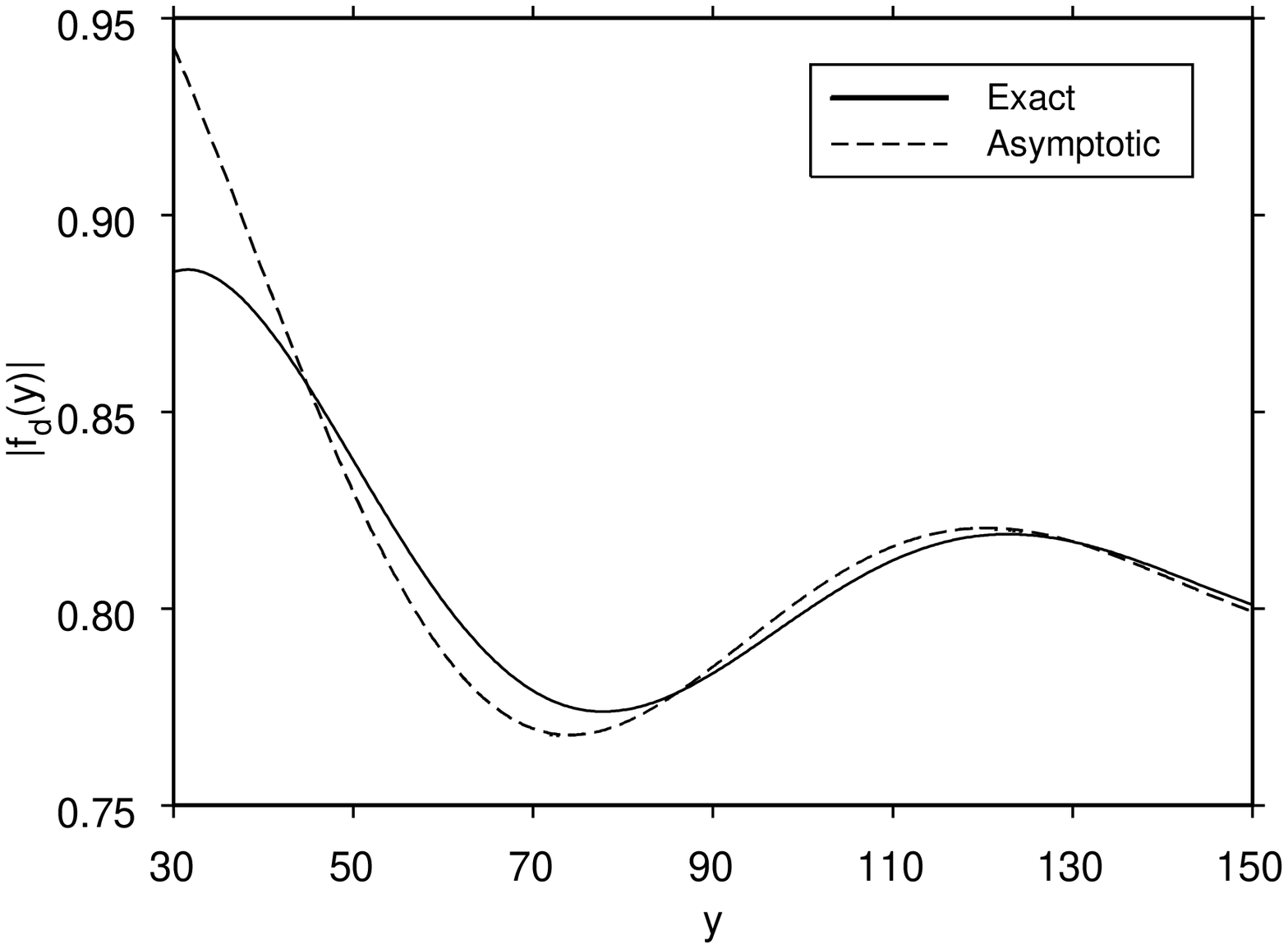}
\end{center}
\end{figure}

\pagebreak

\begin{figure}
\begin{center}
{\Large \sf FIGURE 3}
\medskip
\epsfig{file=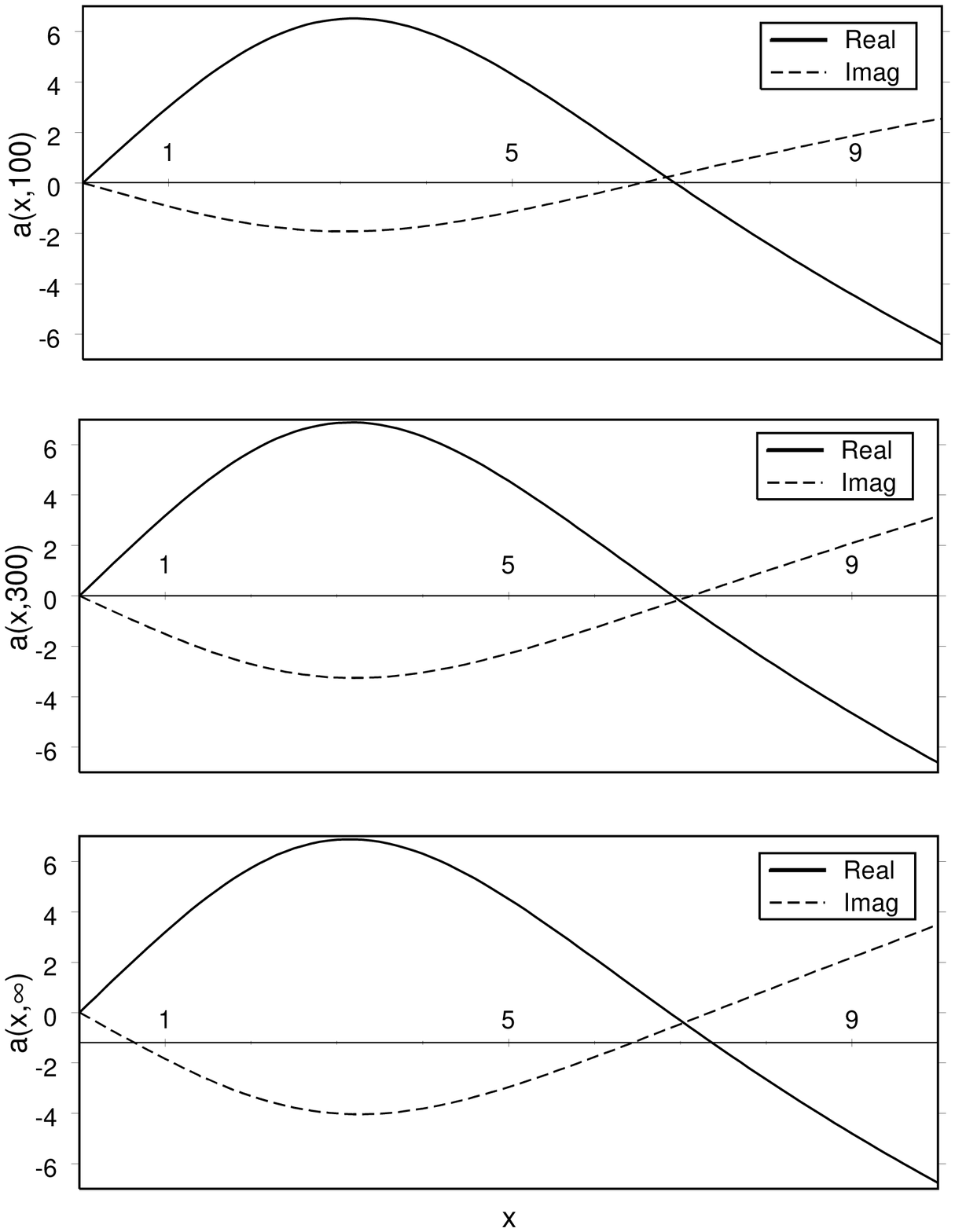}
\end{center}
\end{figure}

\pagebreak

\begin{figure}
\begin{center}
{\Large \sf FIGURE 4}
\medskip
\epsfig{file=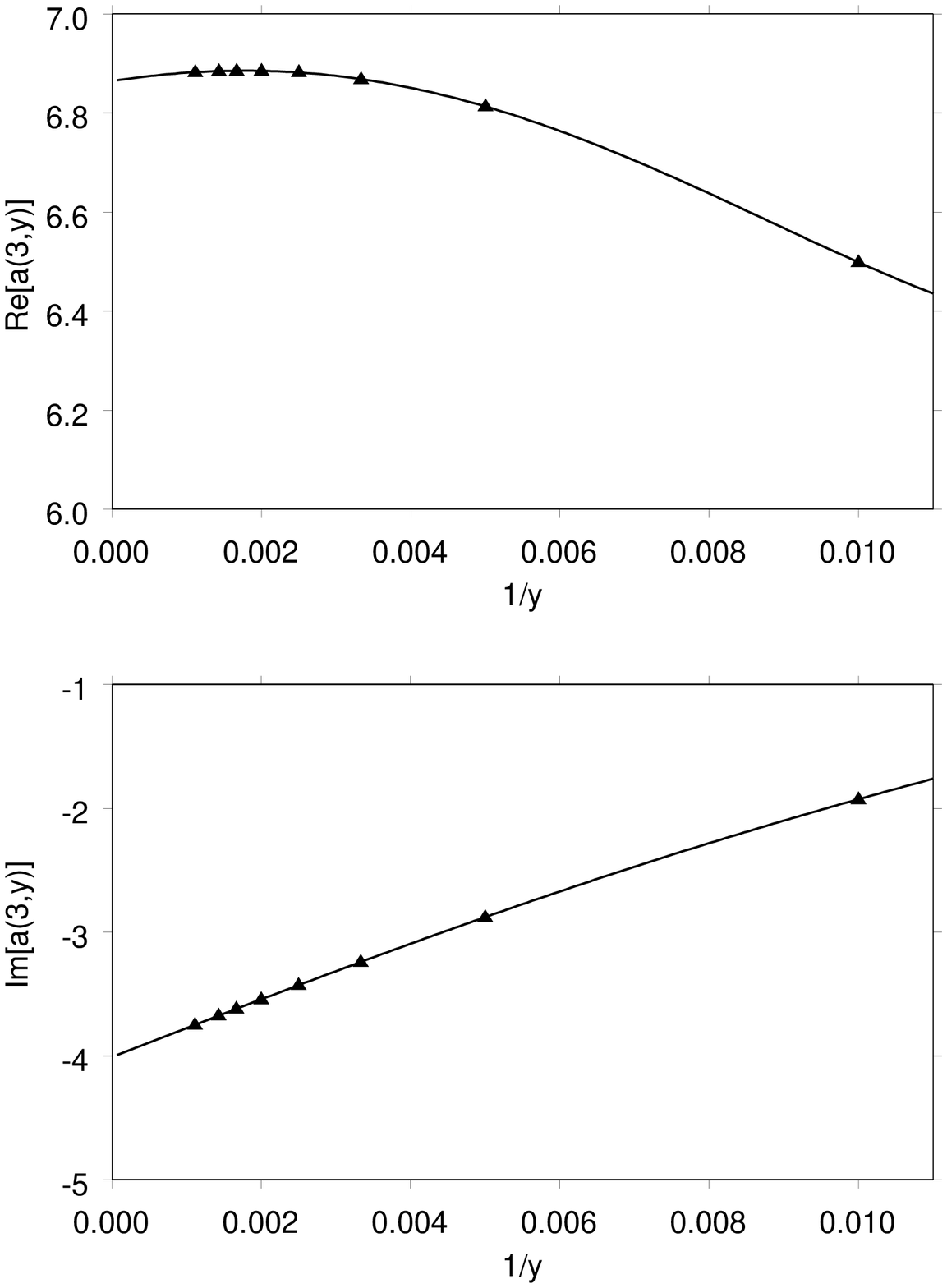}
\end{center}
\end{figure}

\pagebreak

\begin{figure}
\begin{center}
{\Large \sf FIGURE 5}
\medskip
\epsfig{file=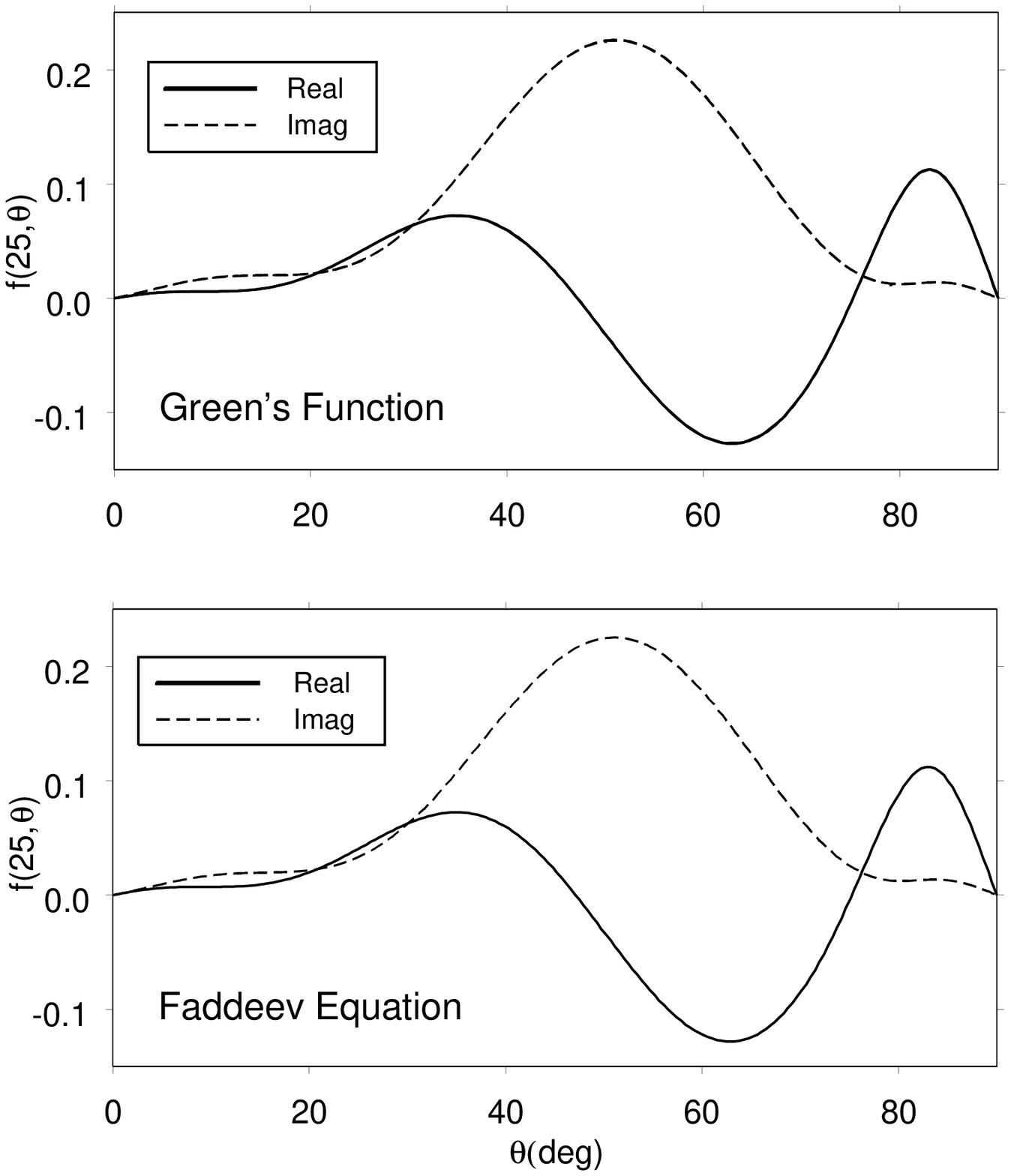}
\end{center}
\end{figure}

\pagebreak

\begin{figure}
\begin{center}
{\Large \sf FIGURE 6}
\medskip
\epsfig{file=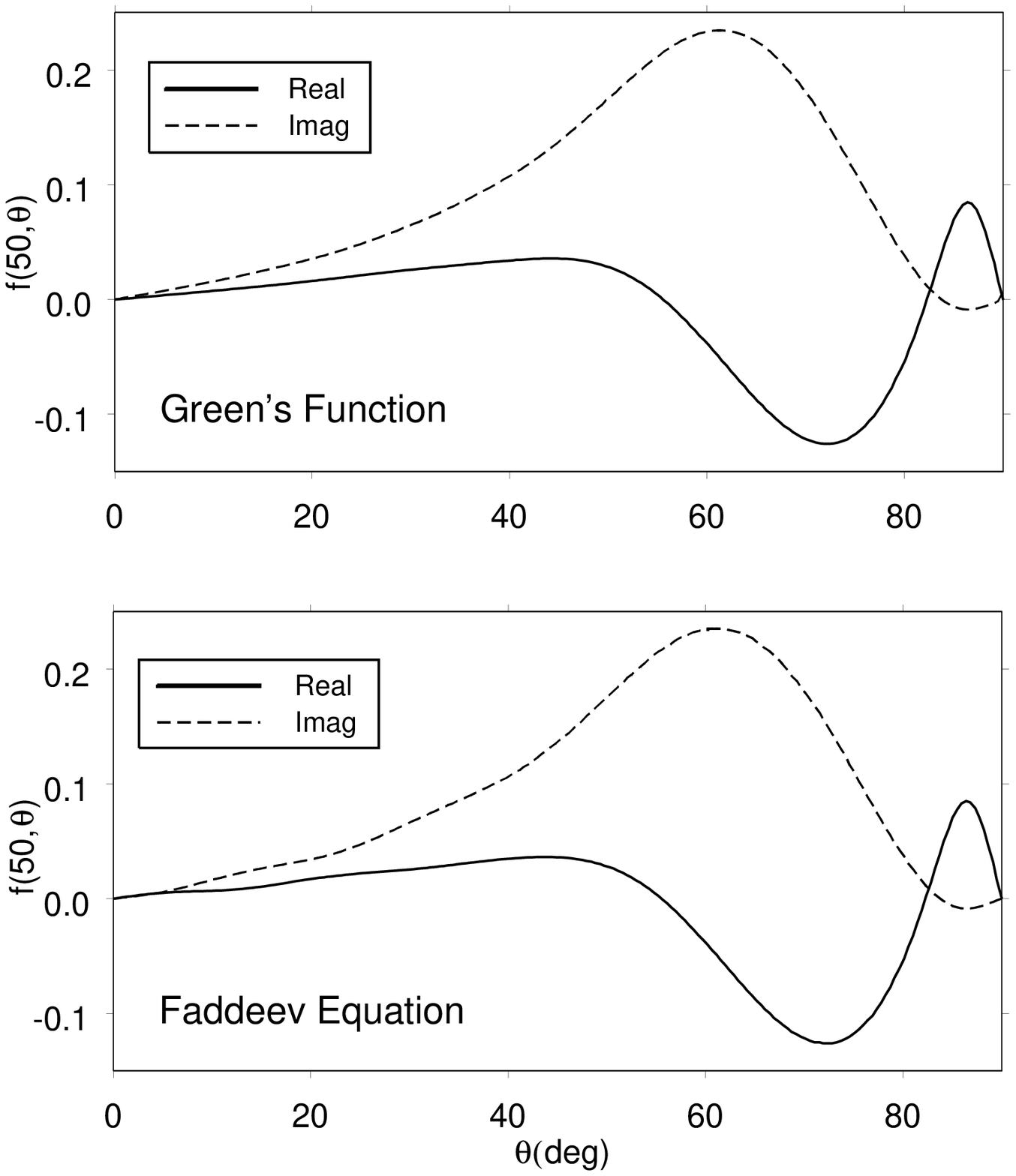}
\end{center}
\end{figure}

\pagebreak

\begin{figure}
\begin{center}
{\Large \sf FIGURE 7}
\medskip
\epsfig{file=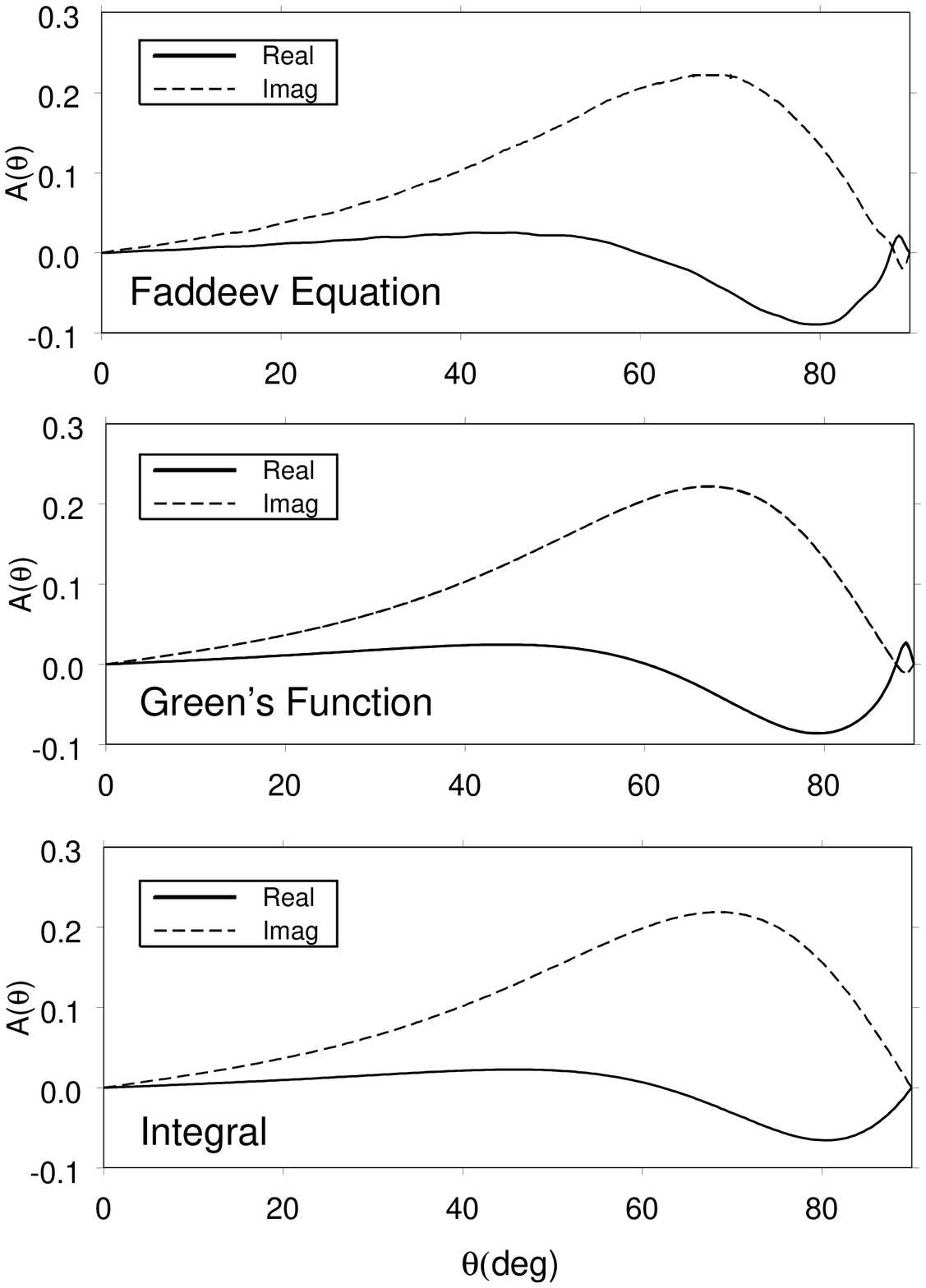}
\end{center}
\end{figure}

\end{document}